\documentstyle[12pt]{article}

\begin{document}
\begin{center}
{\Large\bf To $VMD$, or not to $VMD$, in the quark-gluon plasma}
   \vskip 1.0cm
       {\bf ROBERT D. PISARSKI}\\
       {\it Dept. of Physics, Brookhaven National Laboratory} \\
       {\it Upton, NY, 11973, USA}
   \end{center}
   \vskip 1.5cm
\begin{abstract}
I review results on the shift of the $\rho$ meson mass at nonzero temperature
in a gauged linear sigma model, under the assumption of strict
vector meson dominance.
\end{abstract}
\section{Toy model}

In these proceedings I explain the calculations performed in [1].
The question I address is how the masses of particles, and especially
vector mesons, shift in a hadronic system at
nonzero temperature.  (While the case of nonzero quark density can
be treated by analogous methods, I am {\it not} certain that the results are
qualitatively similar.)  Light vector mesons --- such as the $\rho$, $\omega$,
and $\phi$ --- are of interest because they of their coupling to
low mass dileptons.  For reasons which will become clear, most of my results
only apply to the $\rho$ meson.

Instead of treating vector mesons directly, however, in this section
I consider a toy model which turns out to exemplify much of the important
physics.  Consider a heavy scalar field,
$\widetilde{\Phi}$, coupled to a light scalar field, $\Phi$.  The
$\Phi$ field is an $O(4)$ isovector, and represents a single
$J^P = 0^+$ scalar, the $\sigma$, and an isotriplet of
$J^P = 0^-$ $\vec{\pi}$'s: $\Phi
= (\sigma, \vec{\pi})$.  I assume that the
$\widetilde{\Phi}$ field is very heavy, so that
only fluctuations in the $\Phi$ field matter.

Let the potential terms for the $\widetilde{\Phi}$ and $\Phi$ fields be
\begin{equation}
{\cal V} \; = \frac{\widetilde{m}^2}{2} \; \widetilde{\Phi}^2 \;
+ \; \frac{\kappa}{4}
\; |\Phi|^2 \widetilde{\Phi}^2 \; - \;
\frac{1}{2} \; \mu^2 |\Phi|^2 + \frac{\lambda}{4} \; (|\Phi|^2)^2
\; .
\end{equation}
If the $O(4)$ symmetry is spontaneously broken at zero temperature
by $\mu^2 > 0$, the $\sigma$ field acquires a vacuum expectation
value, $\sigma_0^2 = \mu^2/\lambda$.  After expanding
$\sigma \rightarrow \sigma_0 + \sigma$, the
effects on the $\Phi$ field are standard: the sigma field
becomes massive, $m_\sigma^2 = 2 \lambda \sigma_0^2$,
there is a trilinear coupling $\sigma \vec{\pi}^2$, of
strength $- 2 \lambda \sigma_0 \delta^{a b}$ (the $\delta^{a b}$
is for the isospin of the pions) and a quartic coupling
$\sigma^2 \vec{\pi}^2$, of strength $- 2 \lambda \delta^{a b}$.
(I work with euclidean conventions, so the minus signs are due
to expanding $exp(- S)$, where $S$ is the action, in the path
integral.)

Similarly, the terms in the potential from the interaction between
the two scalar fields is
\begin{equation}
{\cal L} \; = \; \frac{\kappa}{4}
\left(\sigma_0^2 + 2 \sigma_0 \sigma
+ \sigma^2 + \vec{\pi}^2 \right) \widetilde{\Phi}^2 \; .
\end{equation}
After this shift, the mass of the $\widetilde{\Phi}$ field is
\begin{equation}
m^2_{\widetilde{\Phi}}=\widetilde{m}^2
\; + \; \frac{\kappa}{2} \; \sigma_0^2 \; .
\end{equation}
Hence the interaction with the $\Phi$ induces a shift in the
$\widetilde{\Phi}$ mass,
$\delta m^2_{\widetilde{\Phi}}(0)= \kappa \sigma_0^2/2$.
To higher order, there is also
a trilinear coupling $\sigma \widetilde{\Phi}^2$
of strength $- \kappa \sigma_0$, and two quartic couplings,
$\sigma^2 \widetilde{\Phi}^2$, of strength $- \kappa$,
and $\vec{\pi}^2 \widetilde{\Phi}^2$, of strength
$- \kappa \delta^{a b}$.

Having sorted out all the proper conventions of signs and normalization,
it is then straightforward to compute the thermal
shifts in the effective mass of the $\widetilde{\Phi}$ field.
I start with the limit of very low temperatures.  At very low
temperatures, the only fluctuations are those of pions, since fluctuations
in $\sigma$ are suppressed by Boltzman factors of
$exp(-m_\sigma/T)$.  Then there are two diagrams: the quartic
interaction with two pions gives
\begin{equation}
- \widetilde{\Pi}_1 \; = \; ( - \kappa) \; \frac{3}{2} \;
tr \; \frac{1}{K^2} \; = \; - \kappa \; \frac{T^2}{8} \; .
\end{equation}
Here $K^2 = k_0^2 + k^2$, $k_0 = 2 \pi n T$ for a bosonic integral, and
$tr = T \sum_{n= -\infty}^{+\infty} \int d^3k/(2 \pi)^3$.
This is a contribution
to minus the self energy for the $\widetilde{\Phi}$ field
(the minus sign comes from bringing a mass term back into
the action), and so shifts the $\widetilde{\Phi}$ mass
{\it up}.  There is a factor of $3$ from three types of pions,
and a $1/2$ for the symmetry factor of the diagram.

There is a second contribution
from a tadpole diagram involving a $\sigma$ meson propagating
at zero momentum,
\begin{equation}
- \widetilde{\Pi}_2 \; = \; (- \kappa \sigma_0) (- 2 \lambda \sigma_0)
\; \frac{1}{m_\sigma^2} \; \frac{3}{2} \; tr \; \frac{1}{K^2}
\; = \; + \; \frac{3}{2} \; \kappa \; tr \; \frac{1}{K^2} \;
= \; + \kappa \; \frac{T^2}{8} \; .
\end{equation}
Evidently, these two contributions cancel {\it identically},
$\widetilde{\Pi}_1 + \widetilde{\Pi}_2 = 0$.

Thus we have the surprising result that the mass shift
in the $\widetilde{\Phi}$ field vanishes
to $\sim T^2$.  This is the simplest example I know of
of a general result due to Eletsky and Ioffe [2], who argued that
to $\sim T^2$, there is no shift in the pole masses for any
field.  In more complicated examples, such as the vector mesons,
one has to take care to compute the shift in the pole masses,
and not simply the self energies at zero momentum.  In the scalar
example considered here, each diagram is independent of the external
momentum, and so the difference doesn't matter.

One could continue to compute the shift in the thermal
$\widetilde{\Phi}$ mass to higher order.  Including pion
loops order by order in perturbation theory,
such a calculation is an expansion in
$\lambda (1/m_\sigma^2) tr 1/K^2 \sim T^2/\sigma_0^2$.
The term of order $\sim T^4/\sigma_0^2$ is
relatively simple to compute, and I leave it as an exercise to
the reader.  Notice that in this example, the only terms of $\sim T^4$
arise from diagrams at two loop order; there are no terms of $\sim T^2$
at one loop order.

Ultimately, we are not interested simply in the low temperature
regime, but in what happens at the temperature of chiral
symmetry restoration for the $\Phi$ field.  We can easily
compute at this temperature, since then $\sigma_0 = 0$, so we
can work in the chirally symmetric phase.  The calculation is
trivial, because only quartic vertices contribute.  Instead of
just three $\vec{\pi}$'s in the loop, we have three $\vec{\pi}$'s plus one
$\sigma$, so in all the self energy is
\begin{equation}
- \widetilde{\Pi}(T \geq T_\chi) \; = \; - \; \frac{4}{2} \; \kappa \;
tr \, \frac{1}{K^2} \; = - \; \kappa \; \frac{T^2}{6} \; .
\end{equation}
Now I use a trick: for an $O(4)$ field, $T_\chi^2 = 2 \sigma_0^2$,
so I can trade $T_\chi$ for $\sigma_0$.  In this
way, I find that the shift in the $\widetilde{\Phi}$ mass at
$T_\chi$,
$\widetilde{\Pi}(T_\chi) = \delta m^2_{\widetilde{\Phi}}(T_\chi)$
is related to its shift at zero temperature,
$\delta m^2_{\widetilde{\Phi}}(0)= \kappa \sigma_0^2/2$, as
\begin{equation}
\delta m^2_{\widetilde{\Phi}}(T_\chi) \; = \;
\frac{2}{3} \; \delta m^2_{\widetilde{\Phi}}(0) \; .
\label{e4}
\end{equation}
Amusingly enough, this relation is independent of the value of the
coupling constant, $\kappa$.
This is really a trick of the one loop result, and is not general:
to higher loop order, further powers of $\lambda$ enter, and
invalidate this simple form.  Nevertheless, it is notable to find
that at least in weak coupling, the shift in the thermal masses
can be as large as the shifts at zero temperature.  This is
hardly suprising, and so is probably true outside of the weak
coupling regime.

\section{Vector meson dominance}

Going onto vector fields,
I begin by constructing the coupling of vector mesons to the $\Phi$
field in the standard manner, following the assumption of vector
meson dominance ($VMD$) [3].  As I shall show, it turns out that this
assumption is {\it crucial} to the results which follow.  With
vector meson dominance, unique predictions follow; without $VMD$,
there is no unique prediction.

Introducing the matrices $t^0 = {\bf 1}/2$ and
$t^a$, $tr(t^a t^b) = \delta^{a b}/2$, the scalar field $\Phi$ is
\begin{equation}
\Phi = \sigma \, t^0 + i \vec{\pi} \! \cdot \! \vec{t} \; .
\end{equation}
For the left and right handed vector fields I take
\begin{equation}
A^\mu _{l,r} = (\omega^\mu \pm f_1^\mu)t^0
+ (\vec{\rho}^{\, \mu} \pm \vec{a}_1^{\, \mu})\!\cdot\! \vec{t} \; ,
\end{equation}
where $\omega$ and $\vec{\rho}$ are $J^P = 1^-$ fields,
and $f_1$ and $\vec{a}_1$ are $J^P = 1^+$ fields.
$VMD$ [3] tells us to construct an effective lagrangian by
coupling the vector fields to themselves
and to $\Phi$ exclusively through the dimensionless couplings which
follow by promoting the {\it global} chiral symmetry to a
{\it local} symmetry.
Introducing the coupling constant $g$ for vector meson dominance,
the appropriate covariant derivative and field strengths are
$D^\mu \Phi = \partial^\mu \Phi - i g (A^\mu_l \Phi - \Phi A^\mu_r)$
and $F_{l,r}^{\mu \nu}
= \partial^\mu A_{l,r}^\nu - \partial^\nu A_{l,r}^\mu
- i g [A_{l,r}^\mu , A_{l,r}^\nu] $.  The effective lagrangian is then
$$
{\cal L} = tr \left( \left|D^\mu \Phi\right|^2
- \mu^2 |\Phi|^2 + \lambda (|\Phi|^2)^2 - 2 h t^0 \Phi \right. \;
$$
\begin{equation}
\left. \; + \frac{1}{4} (F_l^{\mu \nu})^2 + \frac{1}{4}(F_r^{\mu \nu})^2
+ \frac{m^2}{2} \left( (A_l^\mu)^2 + (A_r^\mu)^2 \right) \right) \; .
\label{e7}
\end{equation}
Besides the parameters $\mu^2$ and $\lambda$ introduced previously,
there is $g$, the coupling for $VMD$, a background field $h$, to
make the pions massive, and a mass term
$\sim m^2$ for the gauge fields.  The essential part of $VMD$ is that
the {\it local} gauge symmetry is only broken by an explicit mass term
for the vector fields.
Much of the physics of this lagrangian can be understood from the kinetic
term for the scalar field,
$$
tr \left( |D^\mu \Phi|^2 \right) =
$$
\begin{equation}
\frac{1}{2} \left( \left(\partial^\mu \sigma + g \vec{a}_1^{\, \mu}
\! \cdot \! \vec{\pi}\right)^2
+ \left( \partial^\mu \vec{\pi} + g \vec{\rho}^{\, \mu}
\!\times\! \vec{\pi}
- g \vec{a}_1^{\, \mu} \sigma \right)^2
+ g^2 \left( \sigma^2 + \vec{\pi}^2 \right) (f_1^\mu)^2 \right)
\label{e2}
\end{equation}
Because it couples to the (isosinglet) current for fermion number,
there are no interactions for the $\omega^\mu$ field.  There
are interactions of $\omega^\mu$ due to effects of the anomaly,
but these are neglected in this work.

$VMD$ really is a peculiar assumption.  At first it sounds most
reasonable: the local gauge symmetry is valid in the limit of
high energies, and so controls the behavior of the dimensionless
coupling constants.  At low energies, the symmetry is broken
by the presence of a mass term for the gauge fields.
But this explanation doesn't really make sense.  One the mass
term is added for the gauge fields, the ultraviolet behavior of
the gauge fields is dramatically altered, so that the
longitudinal terms in the gauge propagator don't fall off as they
should, like $1/P^2$, but only as $(\delta^{\mu \nu} -
P^\mu P^\nu/m^2)/(P^2 + m^2) \sim - (P^\mu P^\nu/P^2)(1/m^2)$.

In this work and elsewhere [1] I assume that $VMD$ works as is
commonly assumed.  An interesting exercise would be to test,
experimentally, precisely how well $VMD$ works.  That is,
parametrize the most general lagrangian consistent with the
global chiral symmetry.  This will involve many more (dimensionless)
coupling constants than $VMD$, which only involves one dimensionless
coupling constant.  Bounds on the deviations of these coupling
constants from $VMD$ could then be obtained by comparison with
the data from, say, dilepton production.

There are many terms which respect the global chiral symmetry,
and yet violate $VMD$.  A simple example is
\begin{equation}
{\cal L}_\xi = \xi \; tr(|\Phi|^2) \;
tr ( (A_l^\mu)^2 + (A_r^\mu)^2 ),
\end{equation}
where $\xi$ is a dimensionless coupling constant.
This term obviously affects the masses for the vector mesons.
Including it, the masses of the $\rho$ and $a_1$ mesons are,
in a phase with $\sigma_0 \neq 0$,
\begin{equation}
m_\rho^2 \; = \; m^2 + \xi \sigma_0^2 \; \; \; , \;\;\;
m_{a_1}^2 \; = \; m^2 + \xi \sigma_0^2 + g^2 \sigma_0^2 \; .
\end{equation}
This illustrates a problem in deciphering terms which respect
$VMD$ from those which do not.  At zero temperature,
for both the $\rho$ and the $a_1$ only the combination
$m^2 + \xi \sigma_0^2$ enters into the masses.  Thus one can't tell,
from that alone, if the mass of the $\rho$ arises from an explicit
mass term, $\sim m$, or dynamically, from a coupling such as
$\sim \xi \sigma_0^2$.  Notice that one still needs the $VMD$ coupling
$\sim g^2$ to split the $a_1$ from the $\rho$.

In principle, the effects of the term $\sim \xi$ could be
determined indirectly.  Including the effects of the ${\cal L}_\xi$,
the current which couples to the photon is
\begin{equation}
j^\mu \; = \; \frac{m^2 + \xi |\Phi|^2}{g^2} \; \rho^\mu_3 \; .
\end{equation}
The term $\sim m^2$ is the standard term of $VMD$; the second,
$\sim \xi$, is new.  Now $|\Phi|^2 = \sigma_0^2 + \ldots$,
so to lowest order its like the masses: only the combination
$m^2 + \xi \sigma_0^2$ enters, and only can't tell $m^2$
from $\xi \sigma_0^2$.  But there is also a term
in the current $\sim \xi \vec{\pi}^2 \rho^\mu$,
which opens up above
the $\rho \pi \pi$ threshold.  A bound could be placed on
$\xi$ from the absence of such an increase in the
dilepton rate above this threshold.  As a multiparticle state
this is a smooth threshold, but it should be measurable (or not)
nevertheless.

\section{$VMD$ at high temperature}

The analysis of gauged linear sigma models at nonzero temperature
is elementary, at least in perturbation theory;
for earlier works, see [4,5].  Of course
one wishes to use the model in a regime of strong coupling, but
perhaps the qualitative features of a weak coupling analysis are
correct.

First I deal with the analysis of the mass shift at the point
of phase transition.  This case is especially simple to
consider, because when $\sigma_0 = 0$, many trilinear
vertices proportional to $\sigma_0$ vanish.  Also, mixing
between $\partial^\mu \pi$ and $\sigma_0 a_1^\mu$, which
is implicit from Eq. (\ref{e2}), vanishes.

A technical comment about the self energies of massive vector
bosons is in order.  For a massive vector boson of mass $m$
and momentum $P$, at tree level the inverse propagator is
\begin{equation}
\Delta^{-1}_{\mu \nu} \; = \;
\delta^{\mu \nu} \left( P^2  + m^2 \right) - P^\mu P^\nu \; .
\end{equation}
For a gauge vector boson, $m=0$, the propagator has a zero
mode, $P^\mu \Delta^{-1}_{\mu \nu} = 0$, which is
eliminated by gauge fixing.  For a massive vector boson,
when $m \neq 0$ there is no zero mode in $\Delta^{-1}_{\mu \nu}$,
and so the propagator is trivially obtained simply by inverting
the inverse propagator.  At tree level,
\begin{equation}
\Delta^{\mu \nu} \; = \;
\left( \delta^{\mu \nu} - \frac{P^\mu P^\nu}{m^2} \right)
\; \frac{1}{P^2 + m^2} \; .
\end{equation}
Going on to include loop effects, at zero temperature the
most general form of the self energy
involves two functions, proportional to
$\delta^{\mu \nu}$ and $P^\mu P^\nu$.  The physical modes of
the propagator can be shown to be uniquely determined by
the coefficient of the inverse propagator proportional to
$\delta^{\mu \nu}$.

At nonzero temperature, in general
things are more complicated.  The self energy involves
four independent functions, $\delta^{\mu \nu}$, $P^\mu P^\nu$,
$n^\mu n^\nu$, and $P^\mu n^\nu + n^\mu P^\nu$, where
$n^\mu = (1,0,0,0)$.  Since without gauge invariance the
vector propagator is immediately invertible, this causes no
complication.  One can compute the general form of the propagator
as a function of these four functions in the self energy.
Of course the coefficient of the propagator
$\sim \delta^{\mu \nu}$ is easiest to compute, since it
is just the inverse of the coefficient of
$\sim \delta^{\mu \nu}$ in the inverse propagator.
The full calculation shows that, assuming no untoward cancellations,
a pole in the term in the propagator $\sim \delta^{\mu \nu}$
is a pole in all of its coefficients.
Consequently, for simplicity sake I limit myself to this term.

There is a second complication of more significance.  In
general, the vector boson self energy, $\Pi(P)$, is a function
of the momentum $P$.  While the limit at zero momentum, $\Pi(0)$,
is of interest for determining the correlation functions at large
distances, the effects upon couplings to dileptons are governed
by the behavior of the propagator on the mass shell.  To lowest
order, this is given by $\Pi(-M^2)$, where $M$ is the mass at tree
level.

With these preliminaries aside, in the symmetric phase, $T \geq T_\chi$,
the self energies of the vector bosons are:
\begin{equation}
\Pi_\rho^{\mu \nu}(P) \; = \;
\Pi_{a_1}^{\mu \nu}(P) \; = \;
\left( \frac{g^2 + \xi}{6}\right) \; T^2 \; \delta^{\mu \nu} + \ldots \; ,
\end{equation}
\begin{equation}
\Pi_\omega^{\mu \nu}(P) \; = \;
\left( \frac{\xi}{6}\right) \; T^2 \; \delta^{\mu \nu} + \ldots \; ,
\label{e3}
\end{equation}
\begin{equation}
\Pi_{f_1}^{\mu \nu}(P) \; = \;
\left( \frac{2 g^2 + \xi}{6}\right) \; T^2 \; \delta^{\mu \nu} + \ldots \; ,
\end{equation}
where the terms neglected are proportional to $P^\mu P^\nu$.

In general, all masses shift with temperature.  There is a single
exception, which may be of experimental significance.  $VMD$ implies
that $g \neq 0$, $\xi = 0$.  In this case, from (\ref{e3}) we see
that the $\omega$ meson mass does {\it not} shift with temperature.
This is due to the form of the covariant coupling for the $\omega$
meson in (\ref{e2}), where the $\omega$ meson decouples from the kinetic
term for the scalar field.

Indeed, one can turn this observation around.  {\it If the $\omega$
meson mass shifts with temperature, then it must be due to terms
which violate vector meson dominance.}  Now in this paper I assume
the $VMD$ holds.  But since the $\omega$ meson, being a narrow
resonance, is relatively simple to pick out of the high multiplicity
environment of ultrarelativistic heavy ion collisions, {\it any}
shift in its mass would be of enormous interest.

A priori, if one is willing to abandon $VMD$, one cannot predict
whether the mass of the $\omega$ goes up, or down, with temperature.
In the limit that
$m^2 = 0$, then the $\rho$ and the $\omega$, and the $a_1$ and
the $f_1$, move together uniformly with temperature.  At the point of phase
transition, all four fields have the same effective mass:
\begin{equation}
m^2_\rho(T_\chi) \; = \;
m^2_{a_1}(T_\chi) \; = \;
m^2_\omega(T_\chi) \; = \;
m^2_{f_1}(T_\chi)|_{m^2 = 0} \; = \;
\frac{2}{3} \; m^2_\rho(0) \; = \; (629 \, MeV)^2 \; .
\label{e5}
\end{equation}
This relation is analogous to (\ref{e4}), where the coupling constant
$\xi$ has been eliminated.

On the other hand, if $m^2 \neq 0$ and $\xi \neq 0$, then the
$\omega$ mass could be greater at $T_\chi$ than at zero temperature.
In this case, there is no simple relation between $m^2_\omega(T_\chi)$
and $m^2_\omega(0)$.

If $VMD$ holds, so $m^2 \neq 0$ and $\xi = 0$, then
$$
m^2_\rho(T_\chi)|_{\xi = 0} = m^2_{a_1}(T_\chi)|_{\xi = 0} =
\frac{1}{3} \left( 2 m^2_\rho(0) + m^2_{a_1}(0) \right)
= (962 \, MeV)^2 \; ,
$$
$$
m^2(\omega)(T_\chi)|_{\xi = 0} = m^2_\omega(0) = (782 \; MeV)^2 \; ,
$$
\begin{equation}
m^2_{f_1}(T_\chi)|_{\xi = 0} =
\frac{1}{3} \left( m^2_\rho(0) + 2 m^2_{a_1}(0) \right)
= (1120 \, MeV)^2 .
\label{e6}
\end{equation}

The dependence of the mass shifts on the choice of lagrangian helps to
relate these results with those of other authors, especially those
of Brown and Rho [6].  Following Georgi [7], Brown and Rho work in
a nonlinear sigma model, so the lagrangian includes
$f_\pi^2 |D_\mu U|^2$ for the nonlinear pion field $U$ and
a mass term for the vector fields as in (\ref{e7}).
Georgi takes $m^2 = \beta f_\pi^2$,
where $\beta$ is a dimensionless coupling constant.
Thus when $f_\pi$ vanishes, as it does for $T \geq T_\chi$, the
masses for the $\rho$, $\omega$, $a_1$, and $f_1$ do as well,
at least at tree level.
My difficulty is that I do not understand why in this effective
theory, the dimensional coupling $m^2$ is directly proportional to
$f_\pi^2$; I would expect the theory to have
{\it two} dimensional coupling constants which vary {\it in}dependently.
Hence I would expect that at $T_\chi$, while $f_\pi = 0$,
$m^2 \neq 0$.

Studies using sum
rules typically find that the mass of the $\rho$ meson decreases by
the time of $T_\chi$ [8] (see, however, [9]).
This can be explained by assuming that $VMD$ is violated by
terms such as $\xi \neq 0$.
{}From the measured phase shifts, Shuryak and Shuryak and Thorsson [10] also
find that the $\rho$ meson mass has decreased by the time of $T_\chi$.

I have adopted $VMD$ [1] because it appears to me to be such a beautiful
and mysterious principle.  Beauty can be deceptively seductive,
however, and
so one should be aware that violations of $VMD$ would lead to phenomenon ---
such as decreasing masses, and shifts in the $\omega$ mass --- that
are forbidden under $VMD$.  Ultimately, experiment will decide.

In that vein,
determining whether or not these shifts in the vector meson peaks
are observable is an involved question.  On the theoretical side,
it is necessary to do a complete analysis, including especially
the effects of thermal broadening of the widths for the $\pi$'s,
$\sigma$'s, and all other fields.  While an analysis in weak coupling
has its obvious limitations, it would be well worth checking that
at least the $\pi$'s have a relatively narrow thermal width.  A
necessary condition for any quasiparticle picture to make sense is
that the quasiparticle width is less than the effective mass of the
field.  Secondly, and more importantly, is to establish that the bulk
of the dileptons are in fact to a phase with temperatures $\sim T_\chi$,
so that one can at least hope to pick out thermal dileptons from the
usual zero temperature backgrounds.

\section{$VMD$ at low temperature}

In this section I conclude by analyzing the shifts in the thermal
masses in the limit of low temperature.  This analysis is mainly
of theoretical interest, and is rather more involved than that at
the point of phase transition.  At the outset, I work in the
chiral limit, $h=0$.

There are three diagrams which contribute to self energy of the
$\rho$ meson at one loop order.
There is a tadpole diagram from the $\rho_\mu^2 \pi^2$ coupling,
\begin{equation}
\delta_1 \Pi^{\mu \nu}_\rho(P) \; = \;
+ 2 g^2 \; \left( \frac{m_{a_1}}{m_\rho} \right)^2 \delta^{\mu \nu}
\; tr \; \frac{1}{K^2} \; ,
\end{equation}
a diagram from $\rho \rightarrow \pi \pi \rightarrow \rho$,
\begin{equation}
\delta_2 \Pi_\rho^{\mu \nu}(P) \; = \;
- g^2 \; tr \; \frac{(2 K + P)^\mu (2 K + P)^\nu}{K^2 (K + P)^2} \; ,
\end{equation}
and, lastly, a diagram from $\rho \rightarrow a_1 \pi \rightarrow \rho$,
$$
\delta_3 \Pi_\rho^{\mu \nu}(P) \; = \;
- 2 g^2 \; (m^2_{a_1} - m^2_\rho)
\left( \frac{m_{a_1}}{m_\rho} \right)^2 \times
$$
\begin{equation}
\; tr \left( \delta^{\mu \nu}
- \frac{(P+K)^\mu (P + K)^\nu}{m^2_{a_1}} \right)
\; \frac{1}{K^2 ( (P+K)^2 + m^2_{a_1} )} \; .
\end{equation}

The analysis is greatly simplified by using two tricks.  The first
is to concentrate on the transverse modes, by extracting only the term
proportional to $\sim \delta^{\mu \nu}$.
The second is to
recognize that we need the value of the self energies on their
mass shell.
At low temperatures, only fluctuations from pions matters, so
the loop momenta $K \sim T$; hence,
since $P \sim m_\rho, m_{a_1} \gg T \sim K$, in the
loop integrals we can expand in powers of $K$, treating $K$ as small
relative to $P$.

In this manner, we see that we can neglect $\delta_2 \Pi_\rho$,
since it cannot produce a term $\sim \delta^{\mu \nu}$ to $\sim T^2$.
In $\delta_3 \Pi_\rho$, we can keep just the term
$\sim \delta^{\mu \nu}$, and approximate
\begin{equation}
\frac{1}{(P+K)^2 + m_{a_1}^2} \sim \frac{1}{P^2 + m_{a_1}^2} \; .
\end{equation}
Thus, to order $\sim T^2$ the term in the $\rho$ self energy
proportional to $\delta^{\mu \nu}$ is
\begin{equation}
\Pi^{\mu \nu}_\rho \sim 2 g^2
\left( \frac{m_{a_1}}{m_\rho} \right)^2
\left( 1 \; - \; \frac{m^2_{a_1} - m^2_\rho}{P^2 + m^2_{a_1}} \right) \;
\delta^{\mu \nu} \; tr \; \frac{1}{K^2} \; .
\end{equation}
This form manifestly vanishes on the $\rho$ mass shell,
$\Pi_\rho(P^2) \sim 0$ at $P^2 = - m_\rho^2$,
as is should by the analysis of Eletsky and Ioffe [2].
It also shows that the self energy does {\it not} vanish off of the
mass shell.  This explicit analysis was verified in the linear model
by Lee, Song, and Yabu [5] and, independently, by myself [1].
Song [5] also demonstrated that it holds in the nonlinear model.

For the $a_1$ meson, there are three diagrams which contribute
at one loop order.
There is a tadpole diagram from the $(a_1^\mu)^2 \pi^2$ coupling,
\begin{equation}
\delta_1 \Pi^{\mu \nu}_{a_1}(P) \; = \;
+ g^2 \left( \frac{m_{a_1}}{m_\rho} \right)^2 \delta^{\mu \nu}
\; tr \; \frac{1}{K^2} \; ,
\end{equation}
a tadpole diagram, involving
the cubic couplings between $a_1^2 \sigma$ and $\sigma \pi^2$,
and the $\sigma$ propagator at zero momentum,
\begin{equation}
\delta_2 \Pi^{\mu \nu}_{a_1}(P) \; = \;
- 3 g^2 \left( \frac{m_{a_1}}{m_\rho} \right)^2 \;
\frac{2 \lambda \sigma_0^2}{m^2_\sigma} \; \delta^{\mu \nu}
\; tr \; \frac{1}{K^2} \; ,
\end{equation}
a diagram from $a_1 \rightarrow \rho \pi \rightarrow a_1$,
$$
\delta_3 \Pi_{a_1}^{\mu \nu}(P) \; = \;
- 2 g^2 (m^2_{a_1} - m^2_\rho)
\left( \frac{m_{a_1}}{m_\rho} \right)^2 \times
$$
\begin{equation}
\; tr \left( \delta^{\mu \nu}
- \frac{((P+K)^\mu (P + K)^\nu}{m^2_\rho} \right)
\; \frac{1}{K^2 ( (P+K)^2 + m^2_\rho )} \; ,
\end{equation}
and, lastly, a diagram from $a_1 \rightarrow \sigma \pi \rightarrow a_1$,
$$
\delta_4 \Pi_{a_1}^{\mu \nu}(P) \; = \;
- g^2 (m^2_{a_1} - m^2_\rho)
\left( \frac{m_{a_1}}{m_\rho} \right)^2 \times
$$
\begin{equation}
\; tr
\left( P^\mu + 2 \left( \frac{m_\rho}{m_{a_1}} \right)^2 K^\mu \right)
\left( P^\nu + 2 \left( \frac{m_\rho}{m_{a_1}} \right)^2 K^\nu \right)
\; \frac{1}{K^2 ( (P+K)^2 + m^2_\sigma )} \; ,
\end{equation}
The calculation of the behavior of these diagrams to leading order
in $\sim T^2$ is as easy as for the $\rho$.  The first two contributions
are independent of momentum; the third reduces in the limit of
large $P$ to a simple form, while the fourth has no term
$\sim \delta^{\mu \nu}$ to this order.  The sum is
\begin{equation}
\Pi^{\mu \nu}_{a_1}(P) \sim
- 2 g^2 \left( \frac{m_{a_1}}{m_\rho} \right)^2
\left( 1 \; + \; \frac{m^2_{a_1} - m^2_\rho}{P^2 + m^2_\rho} \right) \;
\delta^{\mu \nu} \; tr \; \frac{1}{K^2} \; .
\end{equation}
Again, we see that $\Pi_{a_1}(P) \sim 0$ on the $a_1$ mass shell,
$P^2 = - m^2_{a_1}$.

So far I have worked in the chiral limit, $h=0$.  It is trivial
extending the analysis away from the chiral limit.  The only
difference for $h \neq 0$ is that the mass of the $\sigma$ depends
upon $h$ as $2 \lambda \sigma_0^2 + h/\sigma_0$.  Thus there
is no change in the result for the $\rho$ meson: its mass does
not move to $\sim T^2$.  For the $a_1$ self energy, the previous
cancellation is now invalid; $\delta_2 \Pi_{a_1}$ changes, with the
total result
\begin{equation}
m^2_{a_1}(T) \sim  m^2_{a_1}
+ \frac{ g^2 m^2_\pi T^2}{4 m^2_\sigma } + \ldots \; .
\end{equation}
That is, the mass of the $a_1$ {\it increases} with temperature, to
leading order $\sim T^2$ at low temperature.
Of course this result is special to the linear sigma model: it
would not be seen in the nonlinear model, where
$m_\sigma \rightarrow \infty$.

The corrections at next to leading order, $\sim T^4$, are
relatively simple to compute.  Remember that in the toy model,
corrections of $\sim T^4$ only arose from diagrams at two
loop order.  Such corrections are of course present in a gauged
linear sigma model.  But there are also corrections at {\it one}
loop order.  After deriving these
corrections, I show in which regime the terms $\sim T^4$ at one loop
order dominate those to two loop order.

To derive these corrections of $\sim T^4$, terms
$\sim K/P$ are retained.  For example,
\begin{equation}
\frac{1}{(P + K)^2 + m_{a_1}^2} \sim
\frac{1}{P^2 + m^2_{a_1}} \left(
1 - \frac{2 K \cdot P}{P^2 + m_{a_1}^2}
- \frac{K^2}{P^2 + m^2_{a_1}}
+ 4 \frac{(K \cdot P)^2}{(P^2 + m_{a_1}^2)^2}
+ \ldots \right)\; .
\label{e8}
\end{equation}
The term $\sim 1$ on the right hand side gives rise to a contribution
of order $T^2$.  Corrections $\sim K \cdot P$ typically vanish
after integrating over $K$.   Terms $\sim K^2$ contribute to
the integral as $tr K^2/K^2 = tr 1$, which doesn't have a term $\sim T^4$.
The remaining term, $\sim (K \cdot P)^2$, produces an integral
which does contribute a term $\sim T^4$:
\begin{equation}
tr \; \frac{K^\mu K^\nu}{K^2} \; = \;
\frac{\pi^2 T^4}{90} \left( \delta^{\mu \nu} - 4 n^\mu n^\nu \right)
\; .
\end{equation}

Calculating in this manner, it is direct to compute the
shift in the $\rho$ mass:
\begin{equation}
m^2_\rho(T) \sim m^2_\rho - \frac{g^2 \pi^2 T^4}{45 m^2_\rho}
\left(\frac{4 m_{a_1}^2 (3 m_\rho^2 + 4 p^2)}
{(m_{a_1}^2 - m_\rho^2)^2} - 3 \right) + \ldots \; .
\label{e9a}
\end{equation}
In Eqs. (\ref{e9a}) and (\ref{e9b}), all masses on the right hand side
refer to values at zero temperature.
The first term on the right hand side is due to
$\rho \rightarrow a_1 \pi \rightarrow \rho$,
from the piece of $\delta_3 \Pi_\rho$ proportional to
$\delta^{\mu \nu}$;
notice that it is proportional to $1/(m_{a_1}^2 - m_\rho^2)^2$.
The second term on the right hand side is
due to $\delta_2 \Pi_\rho$ and $\delta_3 \Pi_\rho$, from the terms
proportional to $K^\mu K^\nu$.

For the $a_1$, the shift in its thermal mass is found to be:
\begin{equation}
m^2_{a_1}(T) \sim m^2_{a_1} + \frac{g^2 \pi^2 T^4}{45 m_\rho^2} \left(
\frac{4 m_{a_1}^2 (3 m_{a_1}^2 + 4 p^2)}{(m_{a_1}^2 - m_\rho^2)^2}
+ \frac{2 m_\rho^4}{m_{a_1}^2 (m_{a_1}^2 - m_\sigma^2)}
- \frac{m_{a_1}^2}{m_\rho^2} \right) + \ldots \; ,
\label{e9b}
\end{equation}
Similar to the
$\rho$, the first term on the right hand side is due to
$a_1 \rightarrow \rho \pi \rightarrow a_1$, from the term in
$\delta_3 \Pi_{a_1}$ proportional to $\delta^{\mu \nu}$,
and is also proportional to $1/(m_{a_1}^2 - m_\rho^2)^2$.
The second term
arises from $a_1 \rightarrow \sigma \pi \rightarrow a_1$,
$\delta_4 \Pi_{a_1}$,
and so involves the mass of the $\sigma$ meson.  Lastly, the third
term on the right hand side arises from the piece of
$\delta_3 \Pi_{a_1}$ proportional to $K^\mu K^\nu$.

The fascinating feature of these results is that the while the $\rho$
mass goes up, and the $a_1$ mass down by $T_\chi$, they start out in
the opposite direction: the $\rho$ mass goes {\it down} to $\sim T^4$,
and the $a_1$ mass, {\it up}!  Technically, this happens because of
the leading term in each mass, proportional to $1/(m_{a_1}^2 - m_\rho^2)^2$.
For the $\rho$, this factor arises
from a factor of $g^2 \sigma_0^2 = (m_{a_1}^2 - m_\rho^2)$ from the vertices,
times the expansion of
$1/((P+K)^2 + m_{a_1}^2)$ in the denominator.  From (\ref{e8}), the
expansion of this denominator to the relevant order gives a factor
of $(m_{a_1}^2 - m_\rho^2)/(P^2 - m_{a_1}^2)^3$,
so overall, at $P^2 = - m_\rho^2$ the factor is
$- 1/(m_{a_1}^2 - m_\rho^2)^2$.
For the $a_1$, there is again a
factor of $g^2 \sigma_0^2 = (m_{a_1}^2 - m_\rho^2)$ from the vertices,
times the expansion of
$1/((P+K)^2 + m_{\rho}^2)$ in the denominator.  Expanding to the
relevant order gives
of $(m_{a_1}^2 - m_\rho^2)/(P^2 - m_{\rho}^2)^3$,
so at $P^2 = - m_{a_1}^2$ the overall
factor is $+ 1/(m_{a_1}^2 - m_\rho^2)^2$, which
has the opposite sign as the analogous term for the $\rho$.

As of yet, I do not know a deeper explanation for the most peculiar
behavior of the vector meson masses to leading order in low temperature.
I can argue, however, that these results are dominant in the limit of
low temperature, at least in certain regimes.  To show this, remember
$m_{a_1}^2 - m_\rho^2 = g^2 \sigma_0^2 = g^2 \mu^2/\lambda$; thus the
shift in the vector meson masses is, to one loop order, of order
\begin{equation}
\delta m^2|_{one \; loop} \; \sim \; g^2
\; \frac{m_\rho^2}{(m_{a_1}^2-m_\rho^2)^2} \; T^4 \;
\sim \; \frac{\lambda^2}{g^2} \; \frac{m_\rho^2}{\mu^2} \;
\frac{T^4}{\mu^2} \; .
\end{equation}
I must confess that the appearance of a factor of $1/g^2$ is disturbing.
However, taking $\lambda \sim g^2$, overall this term is
$\sim g^2$, which is small; if $\lambda \sim g^4$, $\lambda^2/g^2 \sim g^6$
is even smaller.  This is assuming that the mass scales $m$ and $\mu$
are held fixed.  Since in nature $m_\rho \sim m_\sigma \sim \mu$, this
is reasonable.

Now of course the vector mesons masses will shift not just from diagrams
at one loop order, but from diagrams at two loop order and beyond.
These diagrams involve either higher powers of $\lambda$, or higher
powers of $g^2$.  Consider first diagrams with higher powers of $\lambda$:
one example is a one loop correction to the $\sigma$ propagator on a
one loop tadpole graph.  Such a graph is of order of magnitude
\begin{equation}
\delta m^2|_{two \; loop} \; \sim \; (g^2 \sigma_0) (\lambda \sigma_0)
\; \frac{1}{(m_\sigma^2)^2} \; (\lambda T^2) \; T^2
\; \sim \; \lambda g^2 \; \frac{T^4}{\mu^2} \; ,
\end{equation}

An example of a two loop diagram involving higher powers of $g^2$ comes
from, say, $\rho \rightarrow a_1 \pi \rightarrow a_1 \pi \rightarrow \rho$.
This diagram is
large because it involves two $a_1$ propagators.  In magnitude it is
\begin{equation}
\delta m^2|_{two \; loop} \; \sim \; (g^2 \sigma_0)^2
\; \frac{g^2}{(m_{a_1}^2 - m_\rho^2)^2} \; T^4 \; \sim \;
\lambda g^2 \; \frac{T^4}{\mu^2} \; .
\end{equation}
Thus the ratio of these two loop diagrams to one loop diagrams is
\begin{equation}
\frac{\delta m^2|_{two \; loop}}{\delta m^2|_{one \; loop}}
\; \sim \; \frac{g^4}{\lambda} \; \frac{\mu^2}{m_\rho^2} \; .
\end{equation}
Thus for $\lambda \sim g^2$, the one loop term dominates.  It is certainly
possible to choose $\lambda$ in this range.  If $\lambda \sim g^4$
and $\mu \sim m_\rho$,
the terms at one and two loop order are comparable.
If $m_\rho \gg m_\sigma \sim \mu$,
then the one loop term dominates without restriction on the ratio of
the coupling constants.

What, then, of the opposite limit, when $m_\sigma \gg m_\rho$?
This is the limit which is probed by the nonlinear
$\sigma$ model.  In this case the power counting is slightly different,
because then one wants to keep the vacuum expectation value, $\sigma_0^2
= \mu^2/\lambda$, fixed.  In this limit, the one loop term is
\begin{equation}
\delta m^2|_{one \; loop} \; \sim \;
\frac{1}{g^2} \; \frac{m_\rho^2}{\sigma_0^2} \;
\frac{T^4}{\sigma_0^2} \; ,
\end{equation}
while the two loop term is
\begin{equation}
\delta m^2|_{two \; loop} \; \sim \; g^2 \;
\frac{T^4}{\sigma_0^2} \; ,
\end{equation}
Thus
\begin{equation}
\frac{\delta m^2|_{two \; loop}}{\delta m^2|_{one \; loop}}
\; \sim \; g^4 \; \frac{\sigma_0^2}{m_\rho^2} \; ,
\end{equation}
and for small $g$ the two loop term is much smaller than that at
one loop order.
According to this analysis, the terms $\sim T^4$, Eqs.
(\ref{e9a}) and (\ref{e9b}), would be the same if computed in a
gauged nonlinear sigma model with strict $VMD$, taking of course
the limit $m_\sigma \rightarrow \infty$.

The corrections of $\sim T^4$ have also been analyzed by Eletsky and Ioffe
[11], who find that both masses decrease with temperature.  This
can happen in a nonlinear model in which $\xi \neq 0$, following the
discussion in the previous section.
In this conference, S.-H. Lee [12] argued that the terms $\sim T^4$ are
not uniquely defined.  We agree: there is a unique prediction only under
the assumption of strict $VMD$.

\section{References}

$\;$

[1] R. D. Pisarski, hep-ph/9503328.  See, also, R. D. Pisarski,
hep-ph/9503329 and hep-ph/9503330.

[2] M. Dey, V. L. Eletsky, and B. L. Ioffe, Phys. Lett. {\bf 252B}
(1990) 620;
V. L. Eletsky and B. L. Ioffe, Phys. Rev. {\bf D47} (1993) 3083.

[3] S. Gasiorowicz and D. A. Geffen, Rev. of Mod. Phys. {\bf 41} (1969) 531;
J. J. Sakurai, {\it Currents and Mesons} (Univ. of Chicago Press, Chicago,
1969); U. G. Meissner, Phys. Rep. {\bf 161} (1988) 213.

[4] C. Gale and J. I. Kapusta, Nucl. Phys. {\bf B357} (1991) 65;
C. Song, Phys. Rev. {\bf D48} (1993) 1375;
Phys. Lett. {\bf B329} (1994) 312;
C. Gale and P. Lichard, Phys. Rev. {\bf D49} (1994) 3338;
K. Haglin, nucl-th/9410028, to appear in Nucl. Phys. A.

[5] S.-H. Lee, C. Song, and H. Yabu, hep-ph/9408266;
C. Song, hep-ph/9501364.

[6] G. E. Brown and M. Rho, Phys. Rev. Lett. {\bf 66} (1991) 2720;
Phys. Lett. {\bf B338} (1994) 301; hep-ph/9504250; M. Rho, in
proceedings of this conference;
G. E. Brown, A. D. Jackson, H. A. Bethe, and P. M. Pizzochero
Nucl. Phys. {\bf A560} (1993) 1035;
V. Koch and G. E. Brown, Nucl. Phys. {\bf A560} (1993) 345..

[7] H. Georgi, Nucl. Phys. {\bf B331} (1990) 311.

[8] A. I. Bochkarev and M. E. Shaposhnikov,
Nucl. Phys. {\bf B268} (1986) 220;
H. G. Dosch and S. Narison, Phys. Lett. B {\bf 203} (1988) 155;
R. J. Furnstahl, T. Hatsuda, and S.-H. Lee, Phys. Rev. {\bf D42} (1990) 1744;
C. Adami, T. Hatsuda, and I. Zahed, Phys. Rev. {\bf D43} (1991) 921;
C. Adami and I. Zahed, Phys. Rev. {\bf D45} (1992) 4312;
T. Hatsuda and S.-H. Lee, Phys. Rev. {\bf C46} (1992) 34;
T. Hatsuda, Y. Koike, and S.-H. Lee, Phys. Rev. {\bf D47} (1993) 1225;
T. Hatsuda, Y. Koike, and S.-H. Lee, Nucl. Phys. {\bf B394} (1993) 221.

[9] C. A. Dominguez, M. Loewe, and J. C. Rojas, Z. Phys. {\bf C59} (1993) 63;
C. A. Dominguez and M. Loewe, hep-ph/9406213.

[10] E. V. Shuryak, Nucl. Phys. {\bf A533} (1991) 761;
E. V. Shuryak and V. Thorsson, Nucl. Phys. {\bf A536} (1992) 739.

[11] V. L. Eletsky and B. L. Ioffe, hep-ph/9405371.

[12] S.-H. Lee, in proceedings of this conference.

\end{document}